# Authentication System for Smart Homes Based on ARM7TDMI-S and IRIS-Fingerprint Recognition Technologies

Fredrick R. Ishengoma

*Abstract*—With the rapid advancement in technology, smart homes have become applicable and so the need arise to solve the security challenges that are accompanied with its operation. Passwords and identity cards have been used as traditional authentication mechanisms in home environments, however, the rise of misuse of these mechanisms are proving them to be less reliable. For instance, ID cards can be misplaced, copied or counterfeited and being misused. Conversely, studies have shown that biometrics authentication systems particularly Iris Recognition Technology (IRT) and Fingerprint Recognition Technology (FRT) have the most reliable mechanisms to date providing tremendous accuracy and speed. As the technology becomes less expensive, application of IRT& FRT in smart-homes becomes more reliable and appropriate solution for security challenges. In this paper, we present our approach to design an authentication system for smart homes based on IRT, FRT and ARM7TDMI-S.The system employs two biometrics mechanisms for high reliability whereby initially, system users must enroll their fingerprints and eyes into the camera. Iris and fingerprint biometrics are scanned and the images are stored in the database. In the stage of authentication, FRT and IRT fingerprint scan and analyze points of the user's current input iris and fingerprint and match with the database contents. If one or more captured images do not match with the one in the database, then the system will not give authorization.

## I. INTRODUCTION

SAFETY is one of the most significant necessities for people at home. A smart home refers to a home that is combined with a highly sophisticated automatic systems for monitoring temperature, multimedia, windows, doors, alarms, alerts and various additional tasks monitored by computer systems [1]. A smart home technology offers a remote interface to home automation system itself, through a telephone line, wireless transmission or the internet, monitored through a browser, smartphone or a web browser [2].

There exist several security and authentication mechanisms that can be embedded in a smart home. These include the use of numerical codes like passwords, Personal Identification Number (PIN) and passphrases, security tokens like smart card and biometric authentication methods. However, studies have shown that numerical codes, smart cards and physical keys mechanisms have their associated drawbacks [3,4]. Table one shows the risks associated with using various security mechanisms for smart home authentication.

TABLE I
RISKS ASSOCIATED WITH VARIOUS SECURITY TOKENS

| Token | Risks |
| --- | --- |
| Identity card (ID) | Lost, stolen, duplicate, left at home |
| Physical keys | Lost, stolen, duplicate, left at home |
| Password | Forgotten, shared, observed, hacked |
| Magnetic stripe cards | Lost, stolen, duplicate, left at home |
| Smart card | Lost, stolen, duplicate, left at home |
| Signature | Imitated |

Iris technology uses the iris part of the human eye that features a complex system for authentication. The complex system encompasses the combination of different human eye features [5]. It is considered one of the most reliable authentication mechanisms to date. Fingerprint Recognition Technology (FRT) uses human fingerprint to compare the fingerprint patterns in order to identify a person.

In this paper, we present the design of an authentication system for smart home that combines the two-biometrics mechanisms: IRT and FRT. Our research work aims to define a framework that is most reliable for authentication of smart homes.

We start by providing related work in section 2. Section 3 is dedicated to the overview of biometric authentication. Section 4 presents the proposed authentication system with hardware and software designs. Section 5 describes the implementation of the proposed system. Challenges are discussed in section 6 and section 7 presents the conclusion.

## II. RELATED WORK

To our knowledge, not much work has been done to explore the biometric authentication mechanisms specifically for smart homes. However, it has been noted that, the use of biometrics technology for authentication systems in financial sector has been widely studied with various systems and applications proposed by [6-10].

The work by Rajyalakshmi [11] explores the concept of fingerprint and Iris identification on ATM machines. The proposed system combines a number of authentication mechanisms (id card, RFID, passwords, fingerprint and iris recognition) together with a series of authentication steps for ATM access. While this system seems to have the high level of security, the 5 levels of authentication for the user can be burdensome.

Srinivasan [12] proposed the system that uses height sensors for identification in multi-residents home environment. In this system, ultrasonic range sensors are positioned on top of the door at home where they can be used to recognize residents as they walk throughout a home. However this system is limited since it uses a height, which is a weak biometric. The system is limited to the home environment with a small number of residents where height difference can be used as a factor for resident recognition and cannot be used with large population.

Fahmi [13] proposed an authentication system for smart homes that uses ear recognition. In this system, the home server is connected to user's smartphone to authenticate the user once is within the smart home proximity. The system captures ear image and process its Monogenic Representation Local Binary Pattern (M-LBP). The pattern is then compared to the template in the server for recognition. One limitation of this system is that, the user must have a smartphone that is configures to work with the system every time for authentication. Moreover, it is not known how the system is susceptible to fraud, example when a person walks to the door with a configured smartphone and a picture/3D model of the ear of known system user.

While some of these research works are closely related to our work in terms of biometric authentication mechanism, none of the work has explored the design of the two combined biometric security mechanisms for smart homes environment. This work presents a system for a reliable authentication in smart homes.

### III. BIOMETRIC AUTHENTICATION

Biometric authentication systems are gaining attractiveness as a way of providing access in different environments that needs security. Biometric authentication systems are classified into 2 groups: Physical based mechanisms and behavior based mechanisms.

#### A. Physical Based Mechanisms

Psychological based mechanisms are the ones that emphases on observing the biological and the physiological characters of the human being. Examples of physical features are fingerprint and eye iris.

#### B. Behavior Based Mechanisms

Behavior based mechanisms are the ones that emphases on observing the non-biological or the non-physiological characters of the human being. Examples of behavior-based mechanism include the ones that involve gait and typing patterns biometrics.

Both psychological and behavior based biometrics mechanisms works by comparing the input biometric with the saved biometric template. The comparison provides a matching score using hamming distance, which is used to judge whether the person should be given access or not. Hamming distance is a metric that measures the number of positions between two strings of equal length at which the corresponding symbols are different. Hamming Distance is defined as:

$$Hamming\ Distance = \frac{1}{N} \sum_{J=1}^{N} x_j\ (XOR) y_j$$

Fig. 1 depicts how hamming distance works with the calculated hamming distance between the numbers being 3.

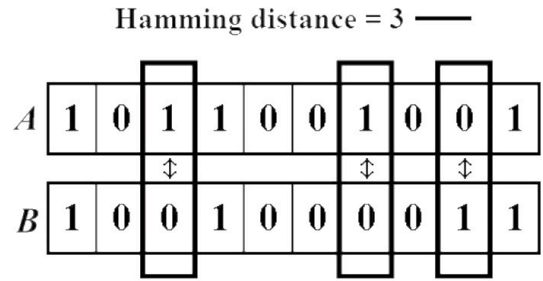

Fig. 1. Example of Hamming Distance

*Algorithm 1:* Typical Biometric-Based Algorithm for Authentication System

```
1: //BT: Biometric
2: //BT: Biometric Template
3: Capture the chosen BT
(Retina/Iris/Fingerprint/Voice/Hand/Face)
4: Process the BT
5: Enroll the BT
6: Store the BT in a local or a central   repository
7: Live-scan the chosen BT
8: Process the BT and extract the BT
9: Compare the scanned BT against the stored
templates
10: Provide a matching score to the application
11: Log the system usage pathway
```

*1) Fingerprint Recognition Technology (FRT)*

This is the type of biometric security that uses the human fingerprint and compares its patterns for identifying a person. The recognition technology involves two steps: Enrolment step and authentication step as shown in Fig. 2. In enrolment step, using fingerprint-capturing device a user's fingerprint image is captured and saved in the database.

In authentication process, the user places his hand on the fingerprint-capturing device whereby it captures his image and compares with the one in the database. If matches then access is granted.

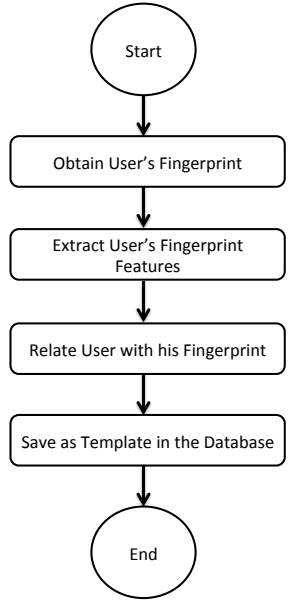

Fig. 2. Fingerprint Authorization Mechanism Stage 1

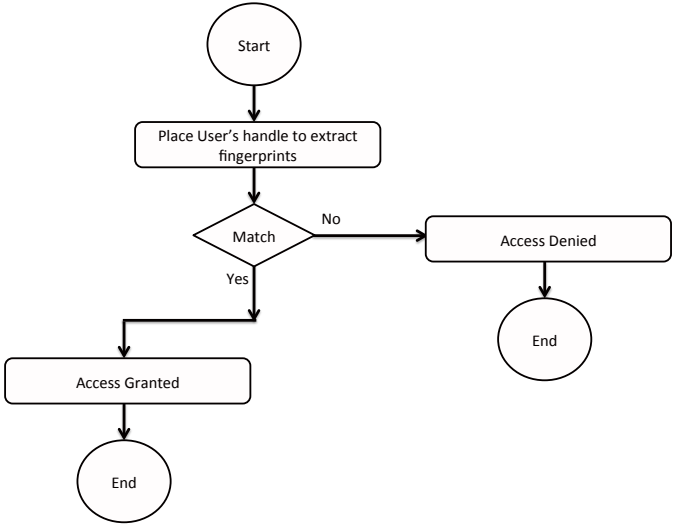

Fig. 3. Fingerprint Authorization Mechanism Stage 2

### 2) Iris Recognition Technology (IRT)

This is the type of biometric mechanism that uses iris of the eye and compares iris patterns for identifying a person. Iris patterns are distinctive and their structures are composed of complicated unique patterns that are not changed till the end of life. With 266 mathematically unique features composed of pigmented vessels and ligaments, the chances of 2 irises to match are very negligible.

The first phase of authentication by using iris technology involves iris image acquisition. User stares at the special camera for some seconds and the camera scan the eye image. The captured digital image is processes by an algorithm and iris is extracted. The iris image is normalized to solve the imaging discrepancies. The patterns that constitute the iris and its surroundings (circular iris, eye lids and pupils) are located by using the formula:

$$max_{(r,x_p,y_0)} \left| G_\sigma * \frac{\Delta}{\Delta x} \oint_{(r,x_p,y_0)} \frac{I(x,y)}{2\pi r} ds \right|$$

Where:

$I(x,y)$ − Eye image, $r$ − Radius, $G_\sigma$ − Gaussian Smoothing Function, $s$ − Contour of the circle given by $r, x_o, y_o$

The circular path is found by changing the radius and the (x, y) positions of the contour of the circle. The precise location of the circular path is obtained through operator application and Gaussian smoothing function. The procedure is the same when searching for eyelids, however here the path is changed from circular to an arc. Normalization is applied to the segmented iris and transformed to polar image coordinates. A Gabor filter 2D form of the image is expressed as:

$$G(x,y) = e^{-\pi \left| \frac{(x-x_0)^2}{\sigma^2} + \frac{(y-y_0)^2}{\beta^2} \right|} e^{-2\pi |u_0(x-x_0)+v_0(y-y_0)|}$$

Where: $(x_0, y_0)$ − Image position, $(\sigma, \beta)$ − (Width, Length), $(u_0, v_0)$ − Modulation

During system authorization process, using hamming distance the two iris codes (user's input and the one stored in the database) are compared. Based on the score authorization decision follows. The overall process is shown in fig. 4.

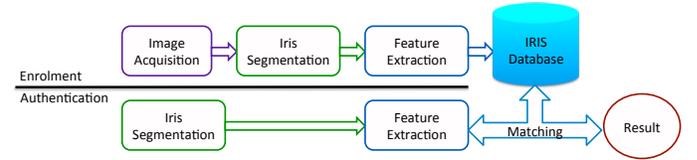

Fig. 4. Iris recognition process

## IV. THE PROPOSED SYSTEM

This section presents the proposed security system that is based on ARM7TDMI-S and the combined biometrics of iris and Fingerprint Recognition Technologies (IRT & FRT).

### A. Hardware Design

#### 1) LPC 2146 Microcontroller

The proposed system is designed to use the LPC 2146 microcontroller that is built on ARM7TDMI-S CPU. The microcontroller is equipped with a 128-bit memory and 32kB – 512kB flash memory. LPC2146 was chosen due its tiny size and low power consumption that makes it more perfect for access control applications.

#### 2) Iris Scanning Camera

This system component is used to capture the iris image of the user. The user will stand, face and stare at this camera. This camera will take the image of user's eye and scan his iris.

#### 3) Fingerprint Scanner (FIM3030HV)

This system component is used to capture the fingerprints of the user. When used at the first time, the user will place his finer on this device, and the device will capture his fingerprint and save in the memory unit. For authentication stage, the user will place his finger, and the device will capture the image and compare with the one saved in the database.

*4) Buzzer*

This system component is used to generate various warning tones example when the user is delaying to place his finger for scanning after the iris recognition step is over.

*5) Output Display*

This system component is used to output result of the authorization process. If the iris patterns of the user and the one in the database matches, then the system will grant the user access and the OD will display a message "Access Granted". However, if the iris patterns of the user and the one in the database does not match, then the system will not grant the user access and the OD will display a message "Access Denied".

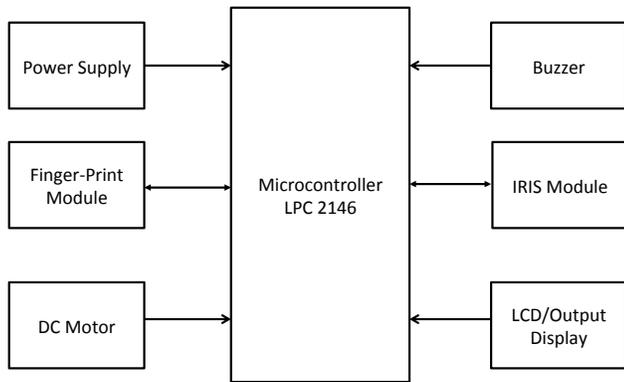

Fig. 5. Block diagram of the proposed system

In this section we present our proposed algorithm that combines fingerprint and iris technology for authentication in smart homes. The proposed algorithm first scans the fingerprint, process and save its image in smart home server. It then scans the iris of the user, process and save its image pattern in the smart home server.

When the user tries to access the system for authorization, the user will have to put the biometric 1 (Fingerprint). The fingerprint will be scanned, processed to obtain its pattern. The obtained pattern will be matched with the one saved in the smart home server using hamming distance. If the patterns match each other, the system will let the user enter the second authorization stage, which is authorization with IRT.

The user will stand in front of the camera whereby his eye will be scanned and processed to obtain its unique patterns. The patterns will be compared with the ones in smart home server, whereby if the patterns match, then the user will be granted access to the smart home. However, if one or both of the authentication mechanism fails the user will be denied access to the smart home.

*Algorithm 2:* Proposed Fingerprint-iris Based Algorithm for Smart Homes Authentication System.
*Input: Biometric (Fingerprint, Iris)*
*Output: System Authorization*

1: //BT1: Fingerprint biometric
2:// BT2: Iris biometric
3: //BT: Biometric Template
4: //BT1: Biometric Template 1
5: //BT2: Biometric Template 2
6: //MS: Matching Score
7: //MSSA: Matching Score for System Authorization
8: Capture user's BT1
9: Process the BT1
10: Enroll the BT1
11: Store the BT1 in a local or a central repository
12: Capture user's BT2
13:Process the BT2
14: Enroll the BT2
15: Store the BT2 in a local or a central repository
16: Live-scan the usersBT1
17:Process the BT1 and extract the BT1
18:Compare the scanned BT1 against the stored templates
19:Provide a matching score to the application
20: If MS for BT1 ≥MSSA
21:    Live-scan the usersBT2
22:    Process the BT2 and extract the BT2
23:    Compare the scanned BT2 against the stored templates
24:    Provide a matching score to the application
25:    If MS for BT2 ≥MSSA
26:        Provide access
27:    else
28:        Deny access
29:    end if
30: else
31: end if
32: Log the system usage pathway

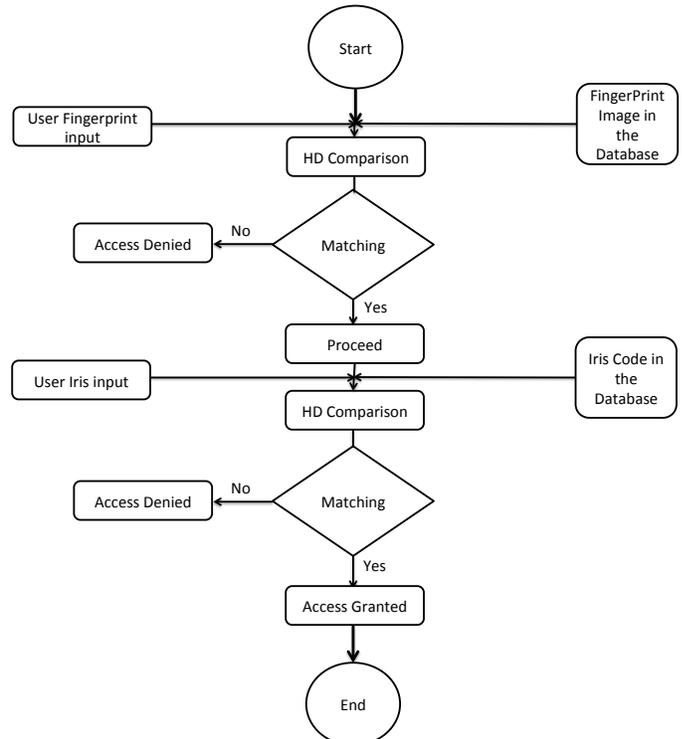

Fig. 6. Flow chart diagram of the prototype

## V. IMPLEMENTATION

The proposed prototype was coded and tested to be working successfully: A CMOS camera integrated with infrared illumination and a cutter filter components was used to capture user's iris image. The infrared illumination had a capacity of up to 790nm. The camera was attached on the door in a fixed position to maintain stability. The FIM3030HV finger print scanner was used.

All the other modules were implemented in the GUI for better and easy use of the system. Fig. 7 shows the CMOS camera used, fig. 8 shows the FIM3030HV module used, fig. 9 shows the welcome screen of the prototype in GUI. Fig. 10 shows the process of iris image processing in the proposed prototype and fi. 11 shows the second stage process where user has to p lace his thumb for finger print recognition.

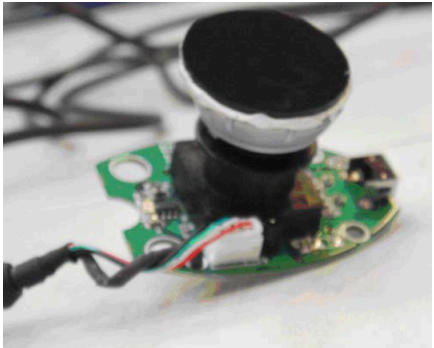
Fig. 7. CMOS Camera

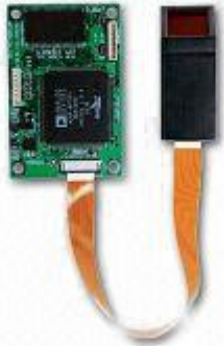
Fig. 8. FIM3030HV Module

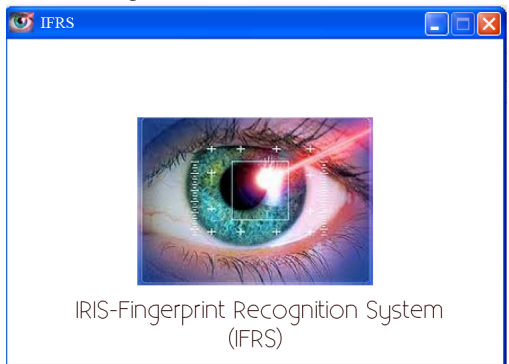
Fig. 9. Welcome screen

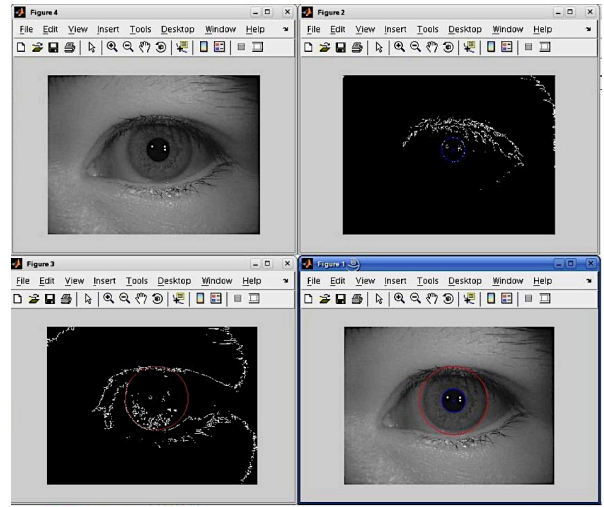
Fig. 10. Iris image processing

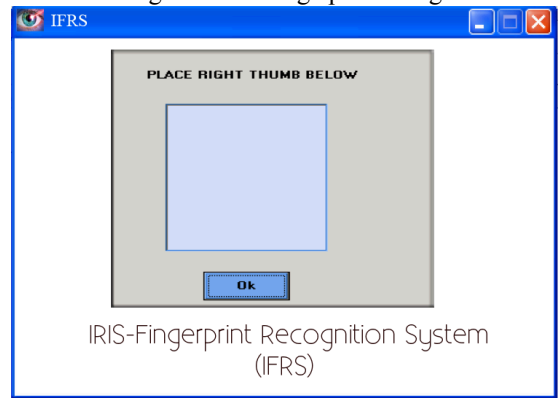
Fig. 11. System waits for user to place his finger

After the iris recognition step, the fingerprint recognition follows. A user places his finger on the FIM3030HV fingerprint-scanner; the scanner scans the fingerprint and compares with it in the database. After the matching procedure, the Output Display will display the result along with authorization or not. The test proved the proposed system to be applicable in real environment.

## VI. CHALLENGES

The proposed system is faced with a challenge for the user who lacks one of the organs that are used in authorization process in the system. Example, if the user lacks both eyes and/or both hands. Also, when the user has contact lenses or sunglasses and also when the user's finger is wet or has wrinkles.

## VII. CONCLUSION

The need for a reliable authentication system at home arises as the misuses of traditional mechanisms (passwords, ID, tokens etc.) increases. This paper proposed a fast and highly reliable authentication system for smart homes that employs double metric mechanisms (Fingerprint and iris recognition technologies). The proposed authentication system scans the user's biometrics and compared with the ones saved in the database for authentication. The result determines if the user will gain access or not. Our future work will focus on dealing with challenges that are facing the proposed prototype example, difficulties when the user has doesn't have an eye/palm, when the user is wearing contact lenses or sunglasses and when the user have wet or wrinkled fingers.

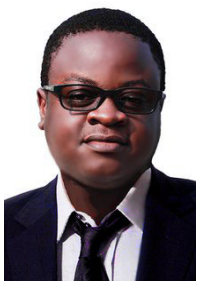

**Fredrick R. Ishengoma** received a Bachelor of Science in Information and Communication Technology Management from Mzumbe University, Tanzania in 2007 and Masters of Engineering in Computer and Information Engineering from Daegu University, South Korea in 2012. He has specialized in the field of Computer Engineering/Information Technology and he has accomplished a number of projects and published a number of research papers in various prestigious international journals with high impact factor. His research interests are Distributed Storage Systems, Hadoop Framework, Block Device Mapping, Erasure Codes, Wireless Sensor Networks and Information and Communication Technology for Development (ICT4D). He is also a member of international professional engineering bodies such as Institute of Electrical and Electronics Engineers (IEEE), International Association of Online Engineers (IAOE), and International Association of Engineers (IAENG).